\documentclass[aps,pra,twocolumn,showpacs,letterpaper,superscriptaddress]{revtex4-1}

\usepackage{graphicx,dcolumn,longtable,epsfig}
\usepackage[usenames]{color}
\usepackage{amssymb}
\usepackage{amsmath}
\usepackage{epstopdf}
\usepackage{bm}
\usepackage{float}
\usepackage{subfigure}

\input epsf.tex
\newcommand{\etal}{{\it et al.}}

\def\jpb#1#2#3{J.~Phys.~B:~At.~Mol.~Phys.~{\bf #1},\ #2\ (#3)}

\def\pr#1#2#3{Phys.~Rev~{\bf #1},\ #2\ (#3)}
\def\pra#1#2#3{Phys.~Rev.~A~{\bf #1},\ #2\ (#3)}

\def\prl#1#2#3{Phys.~Rev.~Lett.~{\bf #1},\ #2\ (#3)}

\def\nat#1#2#3{Nature~{\bf #1},\ #2\ (#3)}

\def\pr#1#2#3{Phys. Rep. ~{\bf #1},\ #2\ (#3)}
\def\yf#1#2#3{Yad. Fiz. ~{\bf #1},\ #2\ (#3)}
\def\sjn#1#2#3{Sov. J. Nucl. Phys. ~{\bf #1},\ #2\ (#3)}
\def\npa#1#2#3{Nucl. Phys. A. ~{\bf #1},\ #2\ (#3)}

\def\etal{{\it et al.}}
\def\bea{\begin{eqnarray}}
\def\eea{\end{eqnarray}}
\def\be{\begin{equation}}
\def\ee{\end{equation}}

\usepackage{color}

\begin{document}
\title{Non-universal bound states of two identical heavy
fermions and one light particle }
\author{ A.~Safavi-Naini}
\email{safavin@mit.edu}
\affiliation{Department of Physics, Massachusetts Institute of Technology, Cambridge, Massachusetts 02139, USA}
\affiliation{ITAMP, Harvard-Smithsonian Center for Astrophysics, Cambridge, Massachusetts 02138, USA}
\author{Seth T. Rittenhouse}
\affiliation{ITAMP, Harvard-Smithsonian Center for Astrophysics, Cambridge, Massachusetts 02138, USA}
\affiliation{Department of Physics and Astronomy, Western Washington University, Bellingham, Washington 98225, USA}
\author{ D. Blume }

\affiliation{ITAMP, Harvard-Smithsonian Center for Astrophysics, Cambridge, Massachusetts 02138, USA}
\affiliation{Department of Physics and Astronomy, Washington State University, Pullman, Washington 99164-2814, USA}
\author{H.~R.~Sadeghpour}
\affiliation{ITAMP, Harvard-Smithsonian Center for Astrophysics, Cambridge, Massachusetts 02138, USA}

\begin{abstract}
We study the behavior of the bound state energy of a system consisting of 
two identical heavy fermions of mass $M$ and a light particle of mass $m$. The heavy fermions interact with the light particle through a short-range two-body potential with positive $s$-wave scattering length $a_s$. We impose a 
short-range boundary condition on the logarithmic 
derivative of the hyperradial 
wavefunction and show that, in the regime where Efimov states are absent, a non-universal three-body state ``cuts through'' the universal three-body states previously described by Kartavtsev and Malykh
[O. I. Kartavtsev and A. V. Malykh, J. Phys. B {\bf{40}}, 1429 (2007)]. 
The presence of the non-universal state alters the behavior of the 
universal states in certain regions of the parameter space. We show that the existence of the non-universal state is predicted accurately by a simple quantum defect theory model that utilizes hyperspherical coordinates. An empirical two-state model is employed to quantify the coupling of the non-universal state to the universal states. 
\end{abstract}

\maketitle
\section{Introduction}
The unprecedented control over ultracold atomic Fermi gases in optical lattices has made them prime candidates for studying and engineering novel quantum phases as well as probing fundamental theories such as the BEC-BCS crossover \cite{Greinter2003,Martin2003,Solomon2003,Hulet2003,Regal2004, Martin2004}. This progress has been facilitated by the tunability of the two-body interactions using Feshbach resonances or by changing the lattice confinement. If the interspecies two-body $s$-wave scattering length $a_s$ is much larger than the range $r_0$ of the underlying two-body potential, the few- and many-body behavior of equal-mass two-component Fermi gases is universal, i.e. completely 
determined by $a_s$. 
 
Presently significant efforts are directed at creating ultracold atomic Fermi gas mixtures composed of two chemically distinct species \cite{Gupta2011, Grimm2010, Martin2012}. This introduces a new parameter, the mass ratio $\kappa$ between the two species. This new parameter affects the many-body physics of the system, allowing one to realize novel quantum phases such as the interior gap superfluid \cite{Wilczek2003}. Here we show that 
at the few-body level this additional degree of freedom leads to new three-body resonances which may destabilize the system, making it harder for experiments to explore novel quantum phases with unequal-mass mixtures. In particular, we study a system of two 
identical heavy fermions with mass $M$, which interact with a light particle through a short-range potential with positive $s$-wave scattering length $a_s$. 

Previous studies revealed two intriguing properties. First,  Kartavtsev and Malykh~\cite{KM} predicted the existence of a universal trimer state with energy $E_{\text{u},1}$ for $\kappa_1<\kappa<\kappa_2$ and the existence of two universal states with energies $E_{\text{u},1}$ and $E_{\text{u},2}$ for $\kappa_2<\kappa<13.606$; $\kappa_1$ and $\kappa_2$ were found to be $8.173$ and $12.917$, respectively. Second, Endo \etal\/~\cite{Endo} investigated how the universal trimer states, which are completely determined by the $s$-wave scattering length and the mass ratio $\kappa$, are connected to Efimov trimers, which have been predicted to exist for $\kappa \gtrsim 13.606$. By analyzing the trimer system within the framework of the Skorniakov-Ter-Martirosian equation with a momentum cutoff 
$\Lambda_c$, Endo \etal\/ predicted the existence of a third class of trimer states, termed crossover trimers, which were shown to continuously connect the universal trimers described by Kartavtsev and Malykh and Efimov trimers.

Our study employs, as in Ref.~\cite{KM}, the hyperspherical coordinates. 
However, while Ref.~\cite{KM} enforced that the hyperradial 
wavefunction vanishes at hyperradius $R=0$, we explore the entirety of physically allowed boundary conditions by introducing a short-range three-body or hyperradial phase $\delta(R_0)$. We determine the eigenspectrum as a function 
of the value of the
three-body phase $\delta$, the hyperradius $R_0$ at which the hyperradial boundary condition is imposed and the mass ratio $\kappa$. 
The universal states of Kartavtsev and Malykh are recovered for $R_0 \to 0$ and $\delta(R_0)=\pi/2$. However, for other boundary conditions we find {\it deviations from universality}, which are linked to the {\it existence of a non-universal three-body state.} Analogous non-universal three-body states have previously been shown to exist \cite{Daily,DailyPRA} (see also Refs.~\cite{Petrov, Nishida, Werner}) in the $a_s\to \infty$ limit. The existence of the non-universal state for positive $a_s$ is described accurately within a quantum defect theory (QDT) framework. Moreover, within a two-state model, deviations from universality are explained as being due to the coupling between the non-universal state and the universal states. 
Our work provides a simple intuitive Schr\"odinger equation 
based description of the energy spectra of heavy-light trimers and 
an alternative means to understanding the 
connection between universal trimers and Efimov trimers. 

The remainder of this paper is organized as follows: 
Section~\ref{sec:framework} describes the hyperspherical framework. 
Section~\ref{sec:numeric} determines the three-body energies as 
functions of $\delta(R_0)$ and $\kappa$ by solving the hyperradial 
Schr\"odinger equation numerically.
Section~\ref{sec:analytic} develops an analytical description, which accounts for the universal and non-universal states of the energy spectrum. Finally, 
Sec.~\ref{sec:conclusion} concludes. 

\section{System Hamiltonian}\label{sec:framework}
We model the interactions between the heavy and light particles by a zero-range two-body pseudopotential with $s$-wave scattering length $a_s$. If we denote the heavy fermions of mass $M$ as $1$ and $2$ and the light particle of mass $m$ as $3$, the Hamiltonian is given by
\be
\label{eq:Htot}
H_{\text{tot}}= 
-\frac{\hbar^2}{2M}(\vec \nabla_1^2+
\vec \nabla_2^2)-\frac{\hbar^2}{2m}\vec\nabla_3^2+V_{\text{int}},
\ee
where $\vec r_i$ is the position vector of the $i$th particle, $\vec \nabla_i^2$ is the Laplacian of the $i$th particle and 
$V_{\text{int}}=V_{\text{ps}}(\vec r_{13})+V_{\text{ps}}(\vec r_{23})$ with
\be
\label{eq:Vij}
V_{\text{ps}}(r_{ij})=\frac{2 \pi \hbar^2 a_s}{\mu_{\text{2b}}} \delta^{(3)} (\vec r_{ij}) \frac{\partial}{\partial r_{ij}} r_{ij}.
\ee
Here $\mu_{\text{2b}}$ is the reduced mass of the heavy-light pair, $\mu_{\text{2b}}=m\frac{\kappa}{\kappa+1}$, and $r_{ij}=\vert \vec r_{ij} \vert = \vert \vec r_i -\vec r_j \vert$. The pseudopotential 
$V_{\text{ps}}$
imposes the Bethe-Peierls boundary condition on the three-body wave function in the limit $r_{ij}\to 0$.

To solve the Schr\"odinger equation for the Hamiltonian 
$H_{\text{tot}}$, we separate off the center of mass motion and write the relative wave function $\Psi$ in terms of the hyperradius $R$ and five hyperangles, collectively denoted by $\Omega$~\cite{KM, Rittenhouse2010}. The hyperrradius $R$, which provides a measure of the overall size of the system, is defined by $\mu R^2=M [(\vec r_1-\vec R_{\text{cm}})^2+(\vec r_2-\vec R_{\text{cm}})^2]+m(\vec r_3-\vec R_{\text{cm}})^2$, 
where $\mu$ is the three-body reduced mass associated with the hyperradius,
$\mu=m \kappa/\sqrt{2\kappa+1}$,
and $\vec{R}_{\text{cm}}$ denotes the center of mass vector.
We expand the relative wavefunction $\Psi$ in terms of a set of weight functions $F_n(R)$ and adiabatic channel functions $\Phi_n(R;\Omega)$, 
which depend parametrically on the hyperradius $R$~\cite{Macek}, 
\begin{equation}
\label{eq:wf} \Psi(R,\Omega)=\sum_n R^{-5/2} F_n(R)\Phi_n(R;\Omega).
\end{equation}
The adiabatic channel functions $\Phi_n(R;\Omega)$ satisfy the hyperangular Schr\"odinger equation at fixed $R$,
\begin{equation}
\label{eq:haSE} \left[\Lambda^2+\frac{2\mu R^2}{\hbar^2}V_{\text{int}} (R, \Omega)\right]\Phi_n(R;\Omega)=\left[s_n^2(R)-4\right]\Phi_n(R;\Omega).
\end{equation}
In Eq.~\eqref{eq:haSE}, 
$\Lambda$ denotes the grand angular momentum operator, 
which accounts for the kinetic energy 
associated with the hyperangles $\Omega$ \cite{Rittenhouse2010}. 
Inserting Eq.~\eqref{eq:wf} into the relative 
Schr\"odinger equation yields a set of coupled 
equations for the weight functions $F_n(R)$.

In the following, we employ the adiabatic approximation, 
which neglects the coupling between the different 
channels~\cite{Rittenhouse2010,Greene1980,Lin1995}. 
This approximation has been shown to provide a 
qualitatively correct description for a number of three-body 
systems~\cite{KM,Esry1996,Esry1996He,Jonsell2002}.  
In this approximation, the hyperradial Schr\"odinger 
equation for the lowest adiabatic channel reads
\be
\label{eq:Hhs} 
-\frac{\hbar^2}{2\mu}
\left[\frac{d^2}{d R^2}-\frac{s_0^2(R)-1/4}{R^2}+Q_{00}(R)\right] 
F_0(R)=EF_0(R),
\ee
where $Q_{00}(R)$ is the diagonal correction
to the adiabatic energies,
\begin{equation}
Q_{00}(R)=
\langle \Phi_0(R;\Omega)\vert \frac{\partial^2}{\partial R^2}\Phi_0(R;\Omega)\rangle.
\end{equation}

\begin{figure}[h]
\begin{center}
\includegraphics[width=0.45\textwidth]{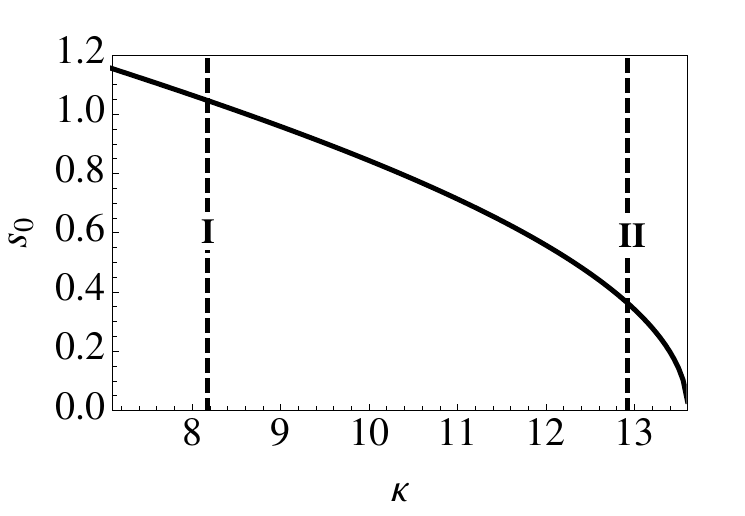}
\caption{The hyperangular eigenvalue $s_0(R_0)$ as a function of the mass ratio $\kappa$ for $R_0/a_s=0.0001$. The dashed lines labeled I and II mark the mass ratios $\kappa_1$ and $\kappa_2$, at which the universal states with energies $E_{\text{u},1}$ and $E_{\text{u},2}$ first become bound. }
\label{fig:s0vsk}
\end{center}
\end{figure}

We determine the three-body energies $E$ using a two step process. First we find the hyperangular eigenvalues $s_0(R)$ and coupling elements $Q_{00}(R)$. Then we solve the radial Schr\"odinger equation, Eq.~\eqref{eq:Hhs}. For states with $L^\pi=1^-$ symmetry and zero-range interactions, the scaled hyperangular eigenvalues $s_0(R)$ can be obtained semi-analytically by solving the 
transcendental 
equation~\cite{KM,Rittenhouse2010}
\begin{widetext}
\be
\frac{R}{a_s}=\frac{(s_0^2-1)(1+2\kappa)^{1/4}\sec(\frac{\pi s_0}{2})\left[s_0 \kappa\,\, {}_2 F_1 \left(\frac{1}{2} (3- s_0), \frac{1}{2}(3+ s_0), \frac{5}{2}, \frac{\kappa^2}{(1+\kappa)^2}\right)-3 (1 + \kappa) \sin(\frac{\pi s_0}{2})\right]}{3 s_0 (1+\kappa)^{3/2}}.
\ee
\end{widetext}
The hyperangular eigenvalues are completely determined by 
$R/a_s$, $\kappa$ and $L^\pi$. 
In the limit $R/a_s \to 0$, 
$s_0$ goes to a constant. 
For the purpose of this study we are only concerned 
with positive values of $s_0^2(R)$ in the small 
$R/a_s$ limit, i.e. we only consider mass ratios 
$\kappa \lesssim 13.606$. 
Figure~\ref{fig:s0vsk} shows the 
scaled eigenvalue $s_0(R_0)$ as a function of $\kappa$ for $R_0/a_s=0.0001$. 
As we discuss in more detail below, 
the value of $s_0(R)$ determines the allowed 
short-range boundary condition of $F_0(R)$. 
In the large $R/a_s$ limit, $Q_{00}$ vanishes and 
the quantity $\frac{\hbar^2(s_0^2(R)-1/4)}{2\mu R^2}$ 
approaches $-E_{\text{2b}}$,
where  $E_{\text{2b}}$ denotes the dimer binding energy,
$E_{\text{2b}}=\hbar^2/(2\mu_{\text{2b}}a_s^2)$.

Equation~\eqref{eq:Hhs} has two linearly independent solutions, 
which scale as $R^{s_0+1/2}$ and $R^{-s_0+1/2}$ in the small $R/a_s$ 
limit and are referred to as the regular and irregular solutions, 
respectively. 
For $s_0(R)\geq 1$ and $R_0/a_s\to0$, 
the irregular solution is not normalizable and thus does not 
contribute, 
implying that $F_0(R)$ goes to zero as $R\to0$~\cite{Petrov,Nishida,Werner}.
On the other hand, in the regime $0<s_0<1$ both solutions 
are well-behaved and must be included when constructing the general 
solution~\cite{Petrov,Nishida,Werner}. 
We parameterize the short-range boundary condition of the wavefunction $F_0(R)$ at $R_0$ using the logarithmic derivative $\mathcal{L}(F_0(R_0))$,
\begin{equation}
\label{eq:logd}R_0 \mathcal{L}(F_0(R_0))=
R_0\frac{\frac{d}{d R} F_0(R)}{F_0(R)}\vert_{R_0}=
\tan (\delta(R_0)),
\end{equation}
where $-\pi/2\leq\delta(R_0) \leq \pi/2$. 
Using this parameterization, we cover all possible short-range phases. 
In the special case of $\delta(R_0)=\pi/2$ and $R/a_s\to 0$, 
this boundary condition and the resulting
three-body energies agree with those of Ref.~\cite{KM}. 
We refer to this boundary condition as ``universal boundary condition'' since the three-body states are completely determined by the regular solution. We refer to the corresponding states as ``universal states'' with energies $E_{\text{u},1}$ for $\kappa_1\leq \kappa\leq \kappa_2$, and $E_{\text{u},1}$ and $E_{\text{u},2}$ for $\kappa_2\leq \kappa\leq 13.606$. In the next section we discuss deviations from universality that occur if $\delta(R_0)$ is not equal to $\pi/2$. These deviations increase with increasing $\kappa$ (for a fixed $R_0$) and increasing $R_0$ (for a fixed $\kappa$).

\section{Numerical Treatment} \label{sec:numeric}
\begin{figure}
\begin{center}
\includegraphics[width=0.45\textwidth]{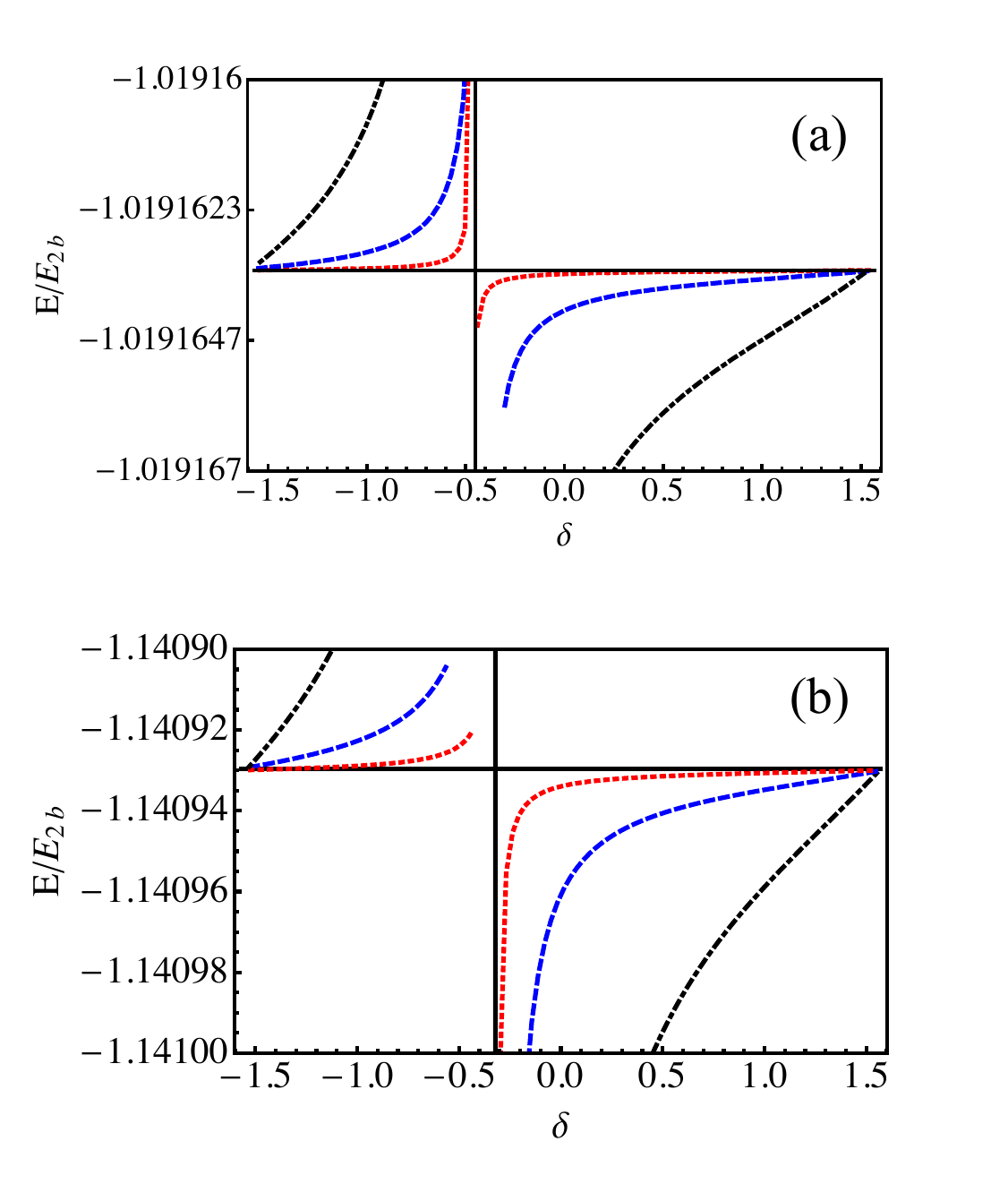}
\caption{(Color online) 
The dotted, dashed, and dash-dotted lines show the scaled three-body 
energy $E/E_{\text{2b}}$ as a function of $\delta$ using 
$R_0/a_s=0.0001$, $0.0003$ and $0.001$, respectively, 
for (a) $\kappa=8.6$ and (b) $\kappa=10$. 
The solid horizontal lines show the energy $E_{\text{u},1}$ 
of the universal state and the solid vertical lines show 
the energy $E_{\text{nu}}$ of the non-universal state 
(see Sec.~\ref{sec:analytic}) for $R_0/a_s=0.001$.} 
\label{fig:Ecomb}
\end{center}
\end{figure}

In this section we examine the behavior of the three-body 
bound states
with $L^{\Pi}=1^-$ symmetry, 
obtained numerically using the shooting algorithm, 
as a function of the logarithmic derivative boundary 
condition for selected mass ratios.
The three-body energies $E$ for $\kappa = 8.6$ and $10$ are shown 
in Figs.~\ref{fig:Ecomb}(a) and (b), respectively. 
Dotted, dashed and dash-dotted lines show $E$ as a function of the three-body phase $\delta$ for $R_0/a_s=0.0001$, $0.0003$ and $0.001$, respectively. 
In Fig.~\ref{fig:Ecomb} the three-body energies have been scaled by the zero-range two-body binding energy $E_{\text{2b}}$. The three-body state becomes unbound with respect to the break-up into a dimer and an atom for $E/E_{\text{2b}}>-1$. For $\kappa=8.6$ and $\delta=\pi/2$, the system supports one three-body bound state. For $R_0/a_s=0.0001$ and $\kappa=8.6$, $E/E_{\text{2b}}$ is nearly constant for $\delta\gtrsim -0.5$. At $\delta\approx -0.5$, referred to as the critical angle $\delta_c(R_0)$, the energy rapidly goes to a large negative value and a second bound state, whose energy is approximately equal to 
$E_{\text{u},1}$, is supported for $\delta\lesssim\delta_c(R_0)$. We refer to the feature in the vicinity of $\delta_c$ as ``three-body resonance''. As $R_0/a_s$ increases  [see dashed and dash-dotted lines in Fig.~\ref{fig:Ecomb}(a)] the width of the three-body resonance increases. Note, however, that the deviations from $E_{\text{u},1}$ are small for all $R_0/a_s$ considered, except for three-body phases very close to $\delta_c(R_0)$. As $\kappa$ increases [see Fig.~\ref{fig:Ecomb}(b)], the overall behavior of the energy spectrum is unchanged. The key trends 
with increasing $\kappa$ are that, at a fixed $R_0$, 
the energy away from $\delta_c(R_0)$ becomes more 
negative and both $\delta_c(R_0)$ and the 
``width'' of the three-body resonance increase (see 
also symbols in Figs.~\ref{fig:deltac} and \ref{fig:beta}). 

\begin{figure}[h]
\begin{center}
\includegraphics[width=0.45\textwidth]{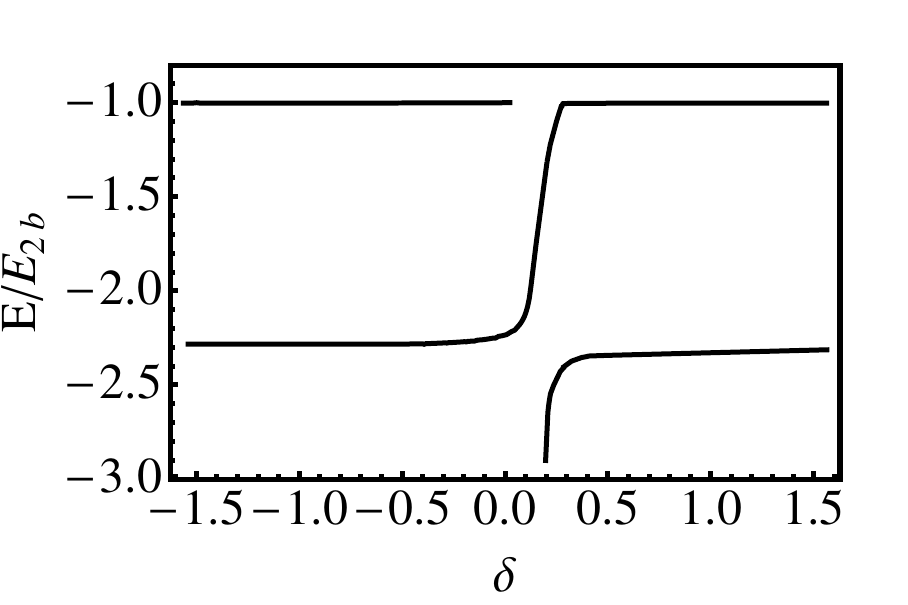}
\caption{The solid lines show the scaled three-body energies $E/E_{\text{2b}}$ as a function of $\delta$ for $\kappa=13$ and $R_0/a_s=0.0001$. The three-body resonance is located at $\delta_c\approx0.15$. }
\label{fig:Ek13}
\end{center}
\end{figure}

For $\kappa\gtrsim \kappa_2$, the three-body system supports a second bound state. 
As an example, Fig.~\ref{fig:Ek13} shows the three-body bound state 
energy for $\kappa=13$ and $R_0/a_s=0.0001$ as a function of $\delta$. 
Away from $\delta_c(R_0)\approx0.15$, there exist 
two bound states whose energies depend weakly on $\delta$. For $\delta\gtrsim\delta_c$, the corresponding hyperradial functions $F_0(R)$ possess $0$ 
nodes and $1$ node, respectively. 
For $\delta\lesssim\delta_c$, 
the corresponding hyperradial functions $F_0(R)$ 
possess $1$ node and $2$ nodes, respectively. This reflects the fact that a new bound state is being pulled in at $\delta\approx \delta_c(R_0)$. 

\begin{figure}[h]
\begin{center}
\includegraphics[width=0.45\textwidth]{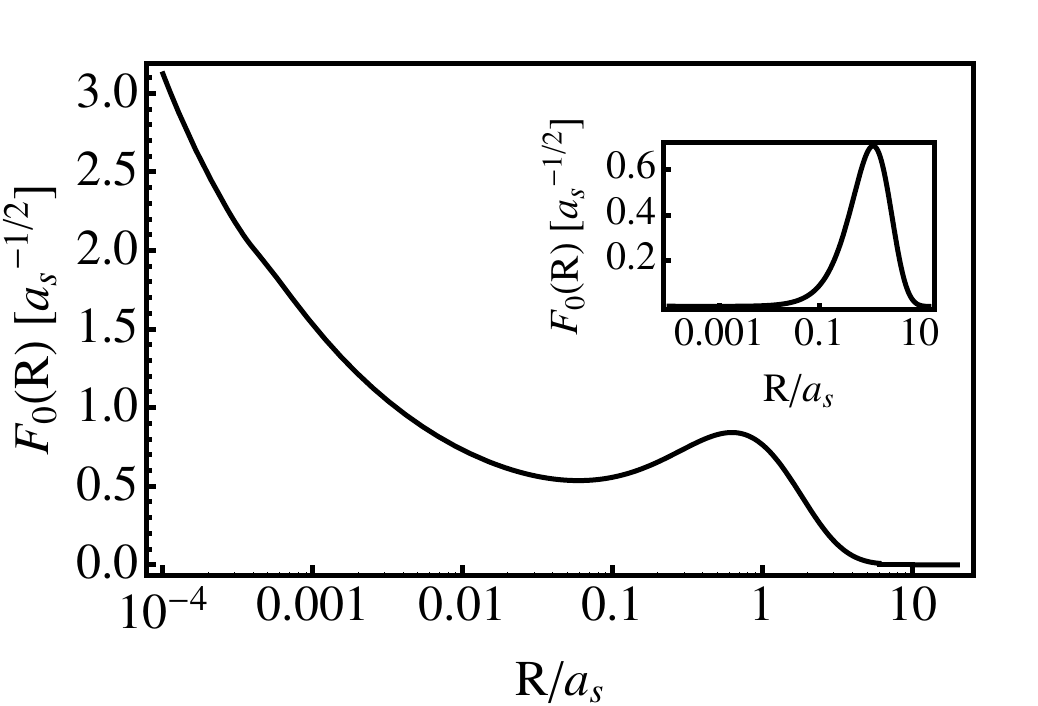}
\caption{The solid line shows the wave function $F_0(R)$ for 
$\kappa=10$, 
$R_0/a_s=0.0001$, $E/E_{\text{2b}}=-1.4$, and $\delta=-0.33$. 
As can be seen in Fig.~\ref{fig:Ecomb}(b), 
this angle is close to $\delta_c$.
For comparison, the inset shows the universal wavefunction 
for $\delta=1.53$ (i.e. away from the three-body resonance).}
\label{fig:k10drop}
\end{center}
\end{figure}

In general, we find that $\delta_c(R_0)$ depends relatively weakly on $R_0$ as long as $R_0/a_s\ll1$. Moreover, 
the three-body energy depends relatively strongly on $R_0$ in the vicinity of $\delta_c$ but comparatively weakly on $R_0$ away from $\delta_c$. This suggests that the states near $\delta_c$ and away from $\delta_c$ can be classified as non-universal and universal, respectively. This interpretation is corroborated by our analysis of the hyperradial wavefunction $F_0(R)$ for $\kappa=10$, $R_0/a_s=0.0001$ and $\delta\approx \delta_c$. The main part of Fig.~\ref{fig:k10drop} shows that the wavefunction has an appreciable amplitude in the small $R/a_s$ region, signaling non-universal behavior. 
Away from $\delta_c$, in contrast, $F_0(R)$ has a vanishingly small amplitude in the
small $R/a_s$ 
region (see inset of Fig.~\ref{fig:k10drop}). We find that $F_0(R)$ depends fairly weakly on $R_0$ when $\delta$ is away from $\delta_c$, lending further support to our assertion that the three-body system behaves universally in this region. For larger $R_0/a_s$, we observe deviations from universality
for a larger range of three-body phases.  Similar non-universal behavior has been previously reported for the three-body system at unitarity~\cite{Daily,DailyPRA} and for the four-body system with positive $s$-wave scattering 
length~\cite{Blume4b}.

\section{Analytical Treatment}\label{sec:analytic}
Section~\ref{sec:analytic}A applies a QDT framework 
to predict the short-range phase $\delta_c(R_0)$ at 
which the non-universal state first becomes bound. 
Section~\ref{sec:analytic}B develops a two-state model to describe the behavior of the universal and non-universal states as a function of $\delta(R_0)$. 
\subsection{QDT Treatment}
To predict the short-range phase $\delta_c(R_0)$ at which the non-universal state first becomes bound with respect to the break-up into a dimer and an atom we apply QDT~\cite{Daily,DailyPRA,Fano,Seaton}. 
In the short-range limit, the wavefunction $F_0(R)$ can be approximated 
by~\cite{Daily,DailyPRA} 
\begin{eqnarray}
\label{eq:SR} F_{\text{SR}}(R)=\nonumber \\
\sqrt{R}[ J_{s_0(R)}(\sqrt{\frac{2\mu E}{\hbar^2}}R)-\tan (\pi\alpha) \,\, Y_{s_0(R)}(\sqrt{\frac{2\mu E}{\hbar^2}} R)],
\end{eqnarray}
where $J_{s_0(R)}$ and $Y_{s_0(R)}$ denote the Bessel functions of the first and second kind, respectively. The quantum defect $\alpha$ controls the relative contribution of the regular solution $J_{s_0(R)}$ and the irregular solution $Y_{s_0(R)}$. A new three-body state is expected to be pulled in when the hyperradial solution is dominated by the irregular solution, i.e. for 
$\alpha=1/2$. The critical angle $\delta_c(R_0)$ is then given by 
\begin{eqnarray}
\tan \delta_c(R_0)=\operatorname{Re}[R_0 \mathcal{L}(F_{\text{SR}}(R_0))],
\end{eqnarray}
where $\operatorname{Re}$ denotes the real part and $F_{\text{SR}}(R_0)$ is 
evaluated for $E=-E_{\text{2b}}$ and $\alpha=1/2$. 

\begin{figure}[h]
\begin{center}
\includegraphics[width=0.45\textwidth]{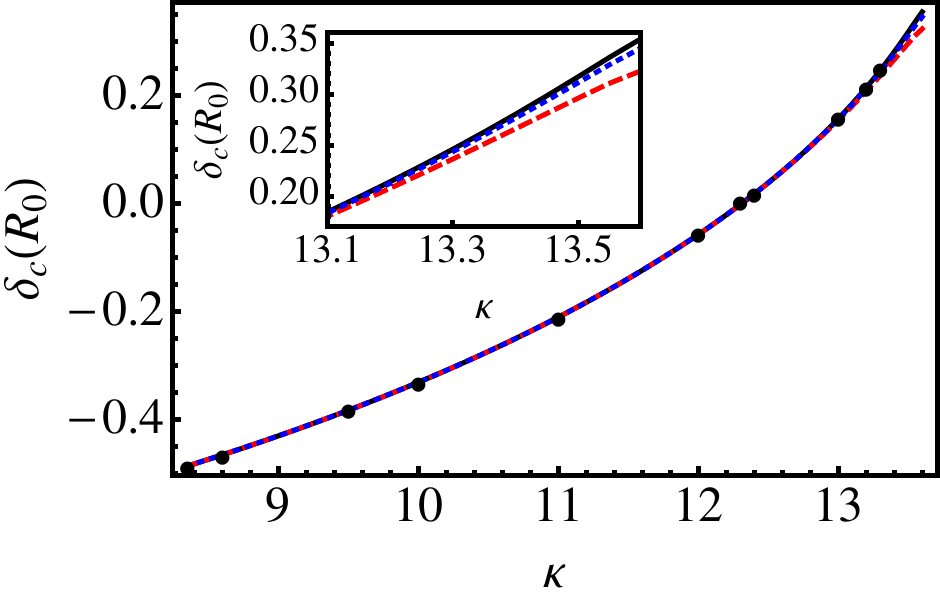}
\caption{(Color online) The critical short-range phase $\delta_c(R_0)$ as a function of $\kappa$. The solid, dotted and dashed lines show the QDT prediction for $R_0/a_s=0.0001, 0.0003$ and $0.001$, respectively.  The circles show $\delta_c(R_0)$ obtained by analyzing the numerical solutions of the hyperradial Schr\"odinger equation, Eq.~\eqref{eq:Hhs}, for $R_0/a_s=0.0001$. The inset shows a blow up of the large $\kappa$ region.}
\label{fig:deltac}
\end{center}
\end{figure}

The solid, dotted and dashed lines in Fig.~\ref{fig:deltac} show 
$\delta_c(R_0)$ determined using the QDT framework for 
$R_0/a_s=0.0001$, $0.0003$ and $0.001$, respectively. 
It is interesting to note that $\delta_c(R_0)$ is very weakly dependent on $R_0$ for small $\kappa$. The symbols in Fig.~\ref{fig:deltac} show the critical angle $\delta_c(R_0)$ obtained by analyzing the numerical solutions of the hyperradial Schr\"odinger equation, Eq.~\eqref{eq:Hhs}, for $R_0/a_s=0.0001$. The numerical results are in excellent agreement with the QDT prediction. For $\kappa\gtrsim13$ (see inset of Fig.~\ref{fig:deltac}), the dependence of $\delta_c$ on $R_0$ becomes more pronounced. 
Figure~\ref{fig:deltac} shows that the three-body system supports a non-universal bound state not only for $\kappa\gtrsim 8.619$, but also for $\kappa\lesssim8.619$ 
[for $\kappa=8.619$ one has $s_0(0)=1$]. 
Thus, it may be surprising at first sight that the system supports 
a non-universal state for $\kappa \lesssim 8.619$
since the irregular solution cannot be normalized if $s_0>1$ for $R_0/a_s\to 0$. However, since we impose the boundary condition at a finite $R_0/a_s$ and not at $R_0/a_s=0$, the resulting wavefunction can, even though it contains an admixture of the irregular solution, be normalized. Correspondingly, non-universal states can exist for $\kappa<8.619$.
In fact, we find that the system supports a non-universal bound state even for $\kappa<\kappa_1$, i.e. for mass ratios where universal states are not supported.

\subsection{Two-state model}
\begin{figure}[h]
\begin{center}
\includegraphics[width=0.5\textwidth]{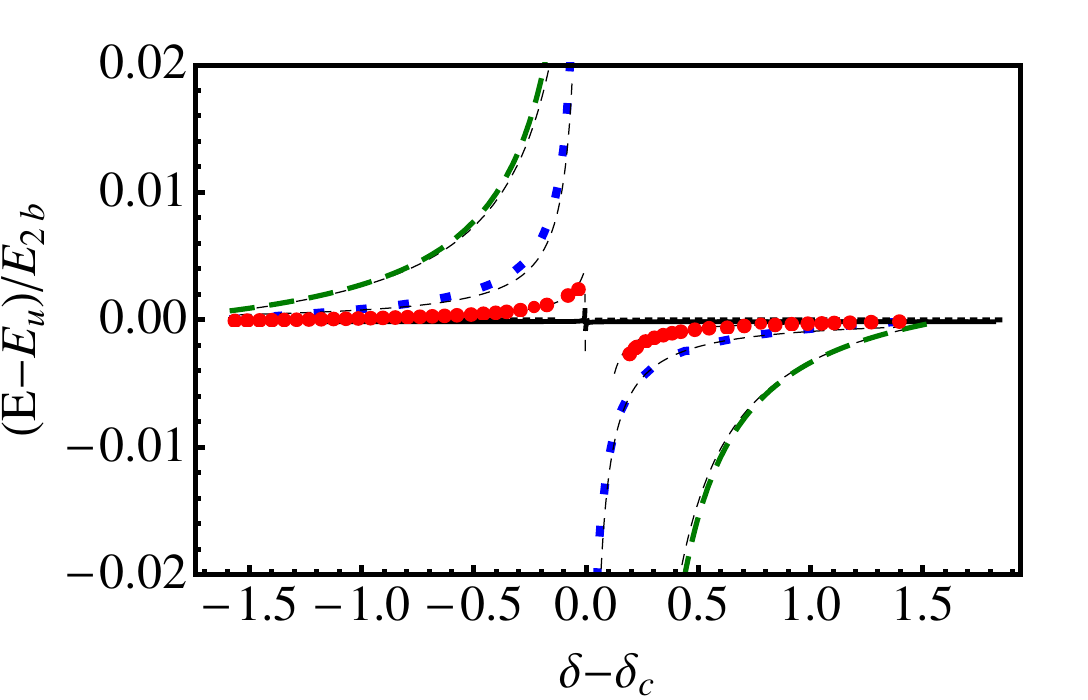}
\caption{(Color online)
Thick solid, dotted and dashed lines show the 
quantity $(E-E_{\text{u},1})/E_{\text{2b}}$  as a function 
of $\delta-\delta_c$ for $R_0/a_s=0.0001$ and $\kappa=10,\, 12.4$, and $13$,
respectively. 
The circles show the quantity $(E-E_{\text{u},2})/E_{\text{2b}}$ 
for $R_0/a_s=0.0001$ and $\kappa=13$. 
The thin dashed lines show the results of the two-state model.}
\label{fig:scaledE}
\end{center}
\end{figure}

In this section we develop a two-state model that describes 
the behavior of the universal and non-universal states 
as a function of $\delta(R_0)$. In our model the three-body resonance is an avoided crossing at $\delta_c(R_0)$ between the universal state with energy 
$E_{\text{u}}$ and a non-universal state with energy $E_{\text{nu}}$. 
In Fig.~\ref{fig:scaledE} we plot the quantity 
$(E-E_{\text{u}})/E_{\text{2b}}$ as a function of $\delta-\delta_c$. 
The thick solid, dotted and dashed lines correspond to 
$\kappa=10, 12.4$ and $13$, respectively, with $R_0/a_s=0.0001$ and 
$E_{\text{u}}=E_{\text{u},1}$. The circles correspond 
to $\kappa=13$, $R_0/a_s=0.0001$ and $E_{\text{u}}=E_{\text{u},2}$.

Motivated by Fig.~\ref{fig:scaledE}, we write the energy of the non-universal state as $E_{\text{nu}} =E_{\text{u}}+\mathcal{A}\frac{\hbar^2}{2\mu R_0^2}(\delta-\delta_c(R_0))$, where $\mathcal{A}$ is a dimensionless scaling constant; in our analysis we use $\mathcal{A}=0.00007$. The term proportional to $\mathcal{A}$ determines the slope at which the non-universal state crosses the universal state. In our parameterization of $E_{\text{nu}}$, $\delta_c(R_0)$ should be interpreted as the three-body phase at which the universal and non-universal states cross. Our numerical results show that this crossing point is nearly identical to the QDT three-body phase at which the non-universal state first becomes bound. Hence we  
use the QDT value in our two-state model. The dimensionless two-state Hamiltonian is then
\begin{equation}
\label{eq:2st} H= \left( \begin{array}{cc}
E_{\text{u}}/E_{\text{2b}} &\beta/E_{\text{2b}}\\
\beta^\star/E_{\text{2b}} &E_{\text{nu}}/E_{\text{2b}} \\
\end{array} \right),
\end{equation}
where $\beta=\beta(R_0, \kappa)$ is the coupling between the universal and non-universal states. 
Within this model, $\beta$ is real and 
the scaled three-body energies $E/E_{\text{2b}}$ 
are given by the eigenvalues of the two-state Hamiltonian. 
We apply the two-state model separately to the universal states with energies $E_{\text{u},1}$ and $E_{\text{u},2}$, yielding separate $\beta$ values for the two universal states. The values of $\beta$ are obtained by fitting the energy spectrum predicted by the two-state Hamiltonian to the numerically determined energies. Thin dashed lines in Fig.~\ref{fig:scaledE} show the results for the two-state model. 
It can be seen that the two-state model provides 
a quantitatively correct description of the
three-body spectra. Moreover, it provides
an intuitive physical picture in which deviations 
from universality arise due to the
coupling of the universal states to the non-universal state.

\begin{figure}[h]
\begin{center}
\includegraphics[width=0.45\textwidth]{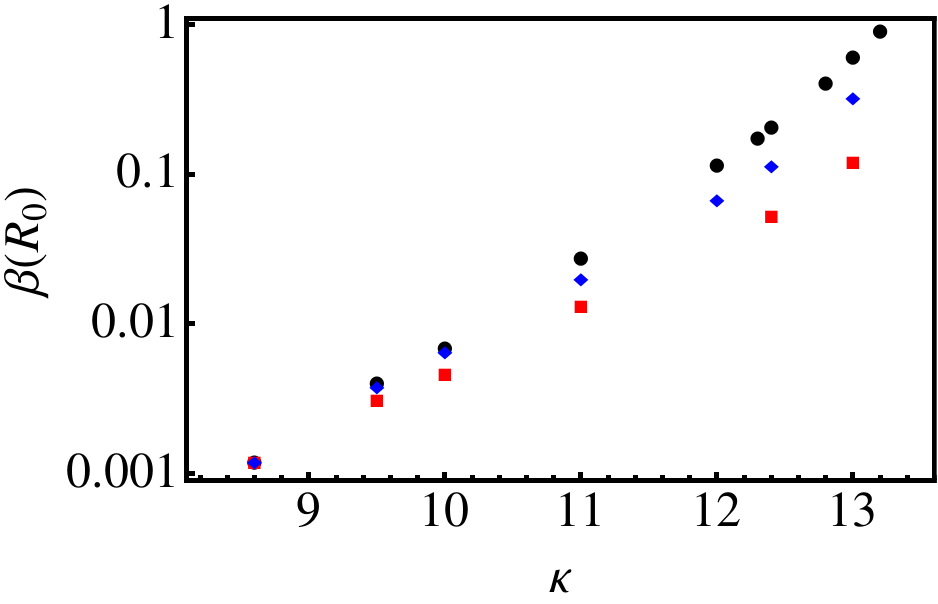}
\caption{(Color online)
Circles, diamonds and squares show the off-diagonal coupling $\beta/E_{\text{2b}}$ used in the two-state model, Eq.~\eqref{eq:2st}, as a function of $\kappa$ for $R_0/a_s=0.0001,0.0003$ and $0.001$, respectively, on a logarithmic scale. We used $E_{\text{u}}=E_{\text{u},1}$ in the two-state model. The dependence of $\beta/E_{\text{2b}}$  on the value of $R_0$ increases with increasing $\kappa$. }
\label{fig:beta}
\end{center}
\end{figure}

The circles, diamonds and squares in Fig.~\ref{fig:beta} 
show $\beta$ for $E_{\text{u}}=E_{\text{u},1}$ and 
$R_0/a_s=0.0001, 0.0003$ and $0.001$, respectively, as a function of $\kappa$. 
The value of $\beta/E_{\text{2b}}$ 
increases with increasing $\kappa$ for fixed $R_0$. 
Moreover, the dependence of $\beta/E_{\text{2b}}$ on the value of $R_0$ increases with increasing $\kappa$. Since $\beta$ determines the coupling between the universal state and the non-universal state, it can be used to quantify the deviations from universality. For mass ratios larger than those considered in this work, the heavy-light trimer system supports three-body bound states with Efimov character \cite{Efimov1970, Efimov1973, Braaten2006}. Within the zero-range framework employed here, the exact number and energy of the Efimov trimers supported depends on the short-range hyperradial boundary condition. If the hyperradial boundary condition is fixed and $\kappa$ is varied, the energy spectrum changes smoothly from (i) deviating from the universal spectrum only in a small region around $\delta_c$ to (ii) deviating from the universal spectrum for a fairly large range of $\delta$ to (iii) supporting three-body states with Efimov character for $\kappa\gtrsim13.606$. Thus, the deviations from universality discussed in this paper can be interpreted as connecting the universal states predicted by Kartavtsev and Malykh and the Efimov trimers.

Our findings are in qualitative agreement with the results reported in Ref.~\cite{Endo}. In that work, the authors imposed a momentum cutoff $\Lambda_c$ in the Skorniakov-Ter-Martirosian equation and used the quantity $\frac{E-E_{\text{2b}}}{E_{\text{u},i}-E_{\text{2b}}}$ ($i=1$ or $2$) 
to determine the ``boundary'' between the universal trimers 
predicted by Kartavtsev and Malykh and the crossover 
trimers in the $(\kappa,(\Lambda_c\, a_s)^{-1})$ parameter space. 
The momentum cutoff was introduced in two ways, using a sharp 
and a Gaussian cutoff. 
We speculate that in our formulation a change in the three-body phase and/or hyperradius $R_0$ corresponds to a change in $\Lambda_c$.  A crossover trimer, in turn, corresponds to a trimer whose energy deviates appreciably from $E_{\text{u},i}$. While the deviations from $E_{\text{u},i}$ are, in our formulation, linked to the coupling of the universal state to a non-universal state, the treatment by Endo \etal\/ does not seem to yield an analogous physical picture. Additionally we speculate that while Ref.~\cite{Endo} employs two different models for the momentum cutoff $\Lambda_c$, it does not explore the entirety of 
the $(R_0, \delta(R_0))$ parameter space. 
In the future it will be interesting to investigate the precise connection between the formulations in the coordinate and momentum spaces by, e.g., comparing the wavefunctions. Such a comparison is needed to check if the above correspondencies are correct.

\section{Conclusion}\label{sec:conclusion}
In this paper we studied a system of two identical heavy 
fermions of mass $M$ and a light particle of mass $m$ with zero-range 
two-body interspecies interactions. 
In particular, we looked at deviations of the three-body bound state 
energies from the universal energies $E_{\text{u},1}$ 
for $\kappa>\kappa_1$ and $E_{\text{u},1}$ and $E_{\text{u},2}$ 
for $\kappa_2<\kappa\lesssim 13.606$ as a function of the 
hyperradial short-range boundary condition. 
We imposed a short-range phase $\delta(R_0)$ using a 
logarithmic derivative boundary condition at various hyperradii $R_0$. 
This parameterization allowed us to explore the full range of 
possible short-range boundary conditions. 

We found that 
(i) for $\delta(R_0)=\pi/2$ the universal states with 
energies $E_{\text{u},1}$ and $E_{\text{u},2}$,
predicted by Kartavtsev and Malykh~\cite{KM}, are recovered; 
(ii) the three-body
states deviate from universality in the vicinity of a 
three-body resonance located at 
the short-range phase $\delta_c(R_0)$, 
at which the non-universal state is first bound; 
(iii) the deviations from universality increase 
with increasing mass ratio $\kappa$ 
(at fixed $R_0$) and with increasing $R_0$ (at fixed $\kappa$); 
(iv) QDT accurately predicts the values of $\delta_c(R_0)$;
(v) a two-state model quantitatively describes the behavior of 
the universal and non-universal states as a function of $\delta(R_0)$;
and, finally, 
(vi) the non-universal bound state exists 
for $\kappa < \kappa_1$ even though universal bound states
are not supported in this regime.

\section{Acknowledgements}
This work was supported by the National
Science Foundation through a grant for the Institute for
Theoretical Atomic, Molecular and Optical Physics at
Harvard University and Smithsonian Astrophysical Observatory.
DB acknowledges support by the NSF through grant PHY-1205443.

\end{document}